\documentclass{iau} 
\usepackage{graphicx}

\title[IAUS 328.~~McIntosh Solar Archive] 
{Beyond sunspots: Studies using the McIntosh Archive of global solar magnetic field patterns}

\author[Sarah E. Gibson, et al.]   
{Sarah E. Gibson$^1$,
David Webb$^2$,
Ian M. Hewins$^2$,
Robert H. McFadden$^2$,
Barbara A. Emery$^2$,
William Denig$^3$,
and Patrick S. McIntosh$^4$}

\affiliation{$^1$High Altitude Observatory, National Center for Atmospheric Research, \\ Boulder, CO, 80301, USA
 \\ email: {\tt sgibson@ucar.edu} \\[\affilskip]
$^2$Institute for Scientific Research, Boston College,\\ Chestnut Hill, MA, USA \\ email: {\tt david.webb@bc.edu} \\[\affilskip]
$^3$National Centers for Environmental Information,\\ National Oceanic and Atmospheric Administration,\\ Boulder, CO, 80305, USA \\ email: {\tt William.Denig@noaa.gov} \\[\affilskip]
$^4$Deceased \\ [\affilskip]
}

\pubyear{2017}
\volume{328}  
\jname{Living Around Active Stars}
\editors{D. Nandi, A. Valio, P. Petit, eds.}
\begin{document}

\maketitle

\begin{abstract}
In 1964 (Solar Cycle 20; SC 20), Patrick McIntosh began creating hand-drawn synoptic maps
of solar magnetic features, based on H$\alpha$ images. These synoptic maps were
unique in that they traced magnetic polarity inversion lines, and connected widely separated
filaments, fibril patterns, and plage corridors to reveal the large-scale organization
of the solar magnetic field. Coronal hole boundaries were later added to the maps,
which were produced, more or less continuously, into 2009 (i.e., the start of SC 24).  The result was a record of
$\sim45$ years ($\sim570$ Carrington rotations), or nearly four complete solar cycles
of synoptic maps. We are currently scanning, digitizing and archiving these maps, with 
the final, searchable versions publicly available at NOAA's National
Centers for Environmental Information. In this paper we present preliminary 
scientific studies using the archived maps from SC 23. 
We show the global evolution of closed magnetic structures (e.g.,
sunspots, plage, and filaments) in relation to open magnetic structures (e.g., coronal holes), and examine how both relate to
the shifting patterns of large-scale positive and negative
polarity regions.
\end{abstract}

\firstsection 
\section{Introduction}

The solar magnetic field is constantly changing, driven by the dynamo below and driving in turn a field that permeates the heliosphere. Concentrated magnetic flux is generated in the Sun's interior and emerges through its surface, e.g., as sunspots. Ongoing diffusion and transport by solar-surface flows results in a shifting pattern of positive and negative magnetic polarity that is an evolving boundary on the global magnetic field.  Because the hot corona results in an expanding solar wind, both ``closed'' and ``open'' magnetic fields extend upwards from this boundary.  Closed-field regions include sunspots, plage, and, if magnetic shear/twist is concentrated at the polarity inversion line (PIL), filaments. Of particular note on a global scale are polar crown filaments, which may extend nearly $360^\circ$ around the sun at high latitudes.  Open-field regions manifest as unipolar coronal holes, which, depending on solar-cycle phase, may appear predominantly at the poles or as isolated structures at lower latitudes.

\begin{figure}[t]
\begin{center}
 \includegraphics[width=5in]{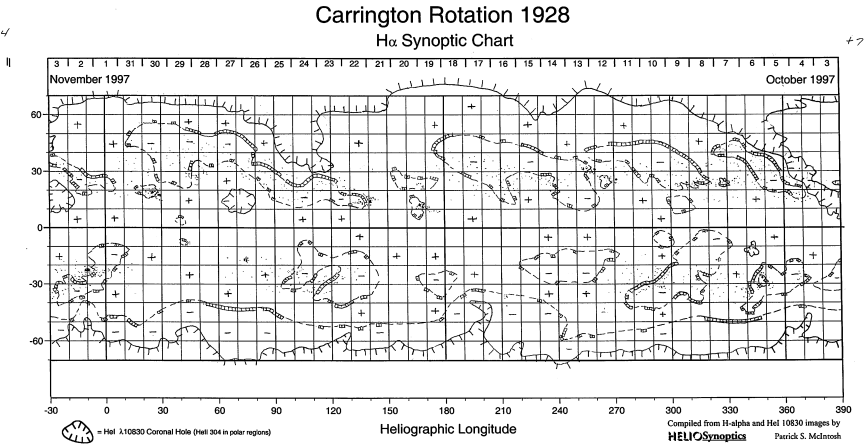} 
 \caption{Example of original, hand-drawn McA synoptic solar map.  Magnetic polarity is indicated by +/-; PILs are dashed lines with filaments indicated by extensions; coronal hole boundaries are indicated by hashed lines; plage by light dots, and sunspots by darker dots.}
   \label{figorigmap}
\end{center}
\end{figure}

\begin{figure}[b!]
\begin{center}
 \includegraphics[width=5in]{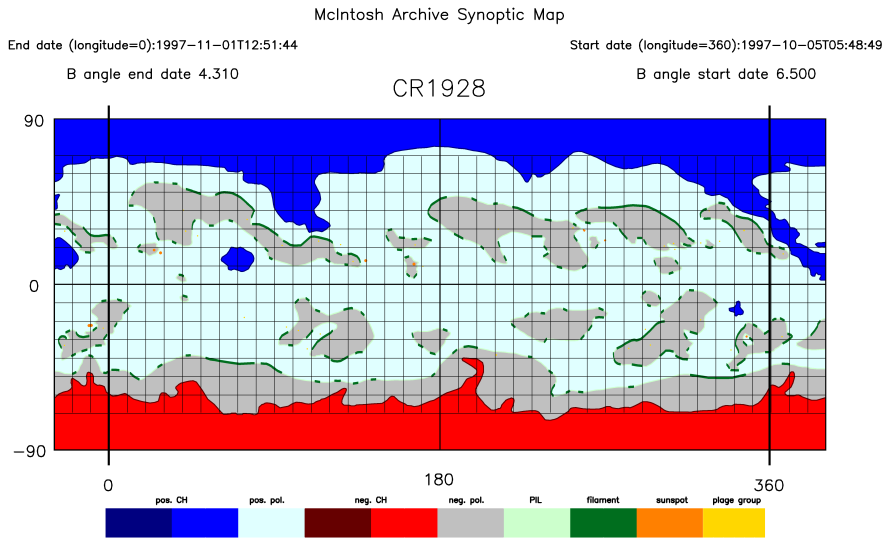} 
 \caption{Example of colorized, digitized McA synoptic solar map  (for the same CROT as Fig. \ref{figorigmap}).  Magnetic features are identified with a distinct color, as described in the legend.}
   \label{figcolormap}
\end{center}
\end{figure}

\section{The McIntosh Archive}

The McIntosh Archive (McA) synoptic maps synthesize these disparate solar magnetic features, open and closed, into a global representation of the evolving solar magnetic field. Over the four decades of their creation, McIntosh consistently used H$\alpha$ daily images to compile the maps. By carefully tracing H$\alpha$ observations of filaments, fibril patterns and plage corridors, McIntosh determined the global structure of the PIL. This H$\alpha$ representation of global magnetism has been shown to correlate well with large-scale magnetic field measured with photospheric magnetograms (\cite[McIntosh, 1979]{mcintosh_79}), and to be particularly useful for tracing the PILs in weak field regions and near the poles of the Sun (\cite[Fox et al., 1998]{foxetal_98}, \cite[McIntosh, 2003]{mcintosh_03}). Starting in 1978, McIntosh added coronal holes to the maps, primarily using ground-based He-I 10830 \AA~ images from NSO-Kitt Peak. Magnetograms were used, when available, to determine the overall dominant polarity of each region. Fig. \ref{figorigmap} shows an example of an original McA map.

\begin{figure}[b!]
\begin{center}
 \includegraphics[width=5.5in]{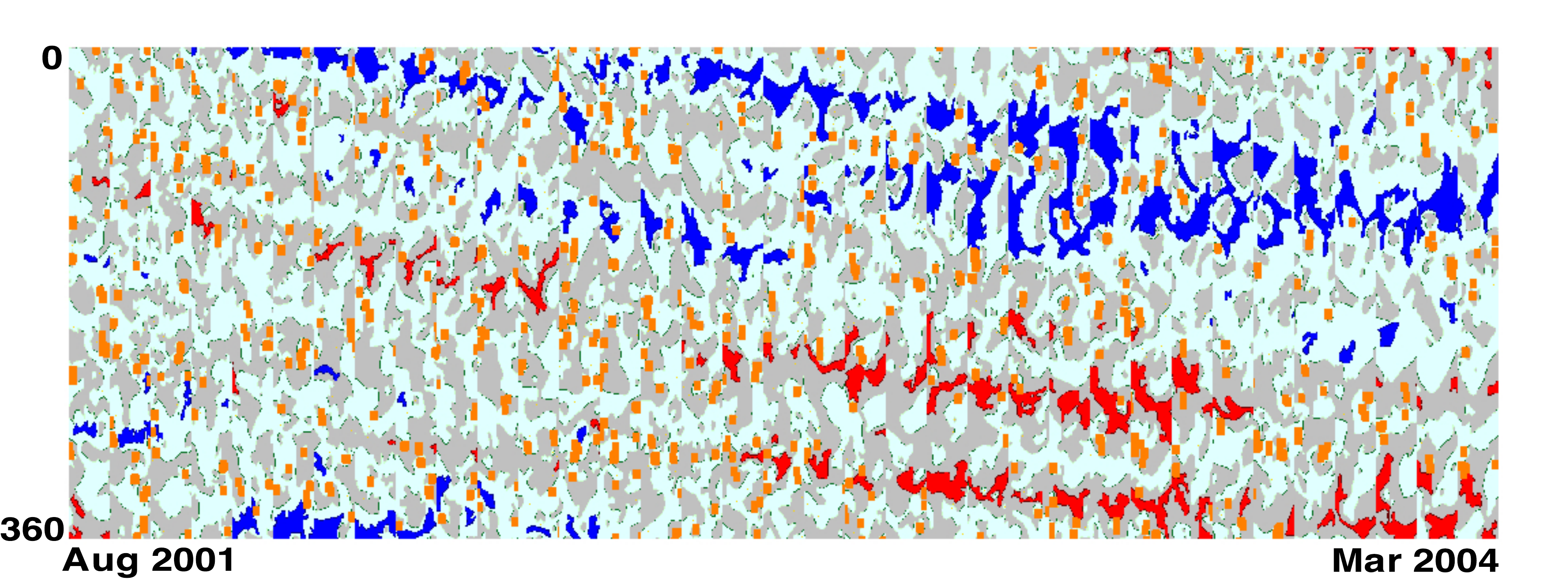} 
 \caption{Equatorial-zone stack plot for CROT 1980 - 2014. Vertical axis is longitude; a sequence of equatorial slices ($-20^\circ$ to $+20^\circ$ latitude) is stacked along the horizontal axis with time increasing to the right. Colors are as in Fig. \ref{figcolormap}, with sunspots (orange), coronal holes (blue=positive, red=negative) and quiet sun (light-blue=positive, grey=negative) indicated.}
   \label{figeq_mid}
\end{center}
\end{figure}

\begin{figure}[t!]
\begin{center}
 \includegraphics[width=5.5in]{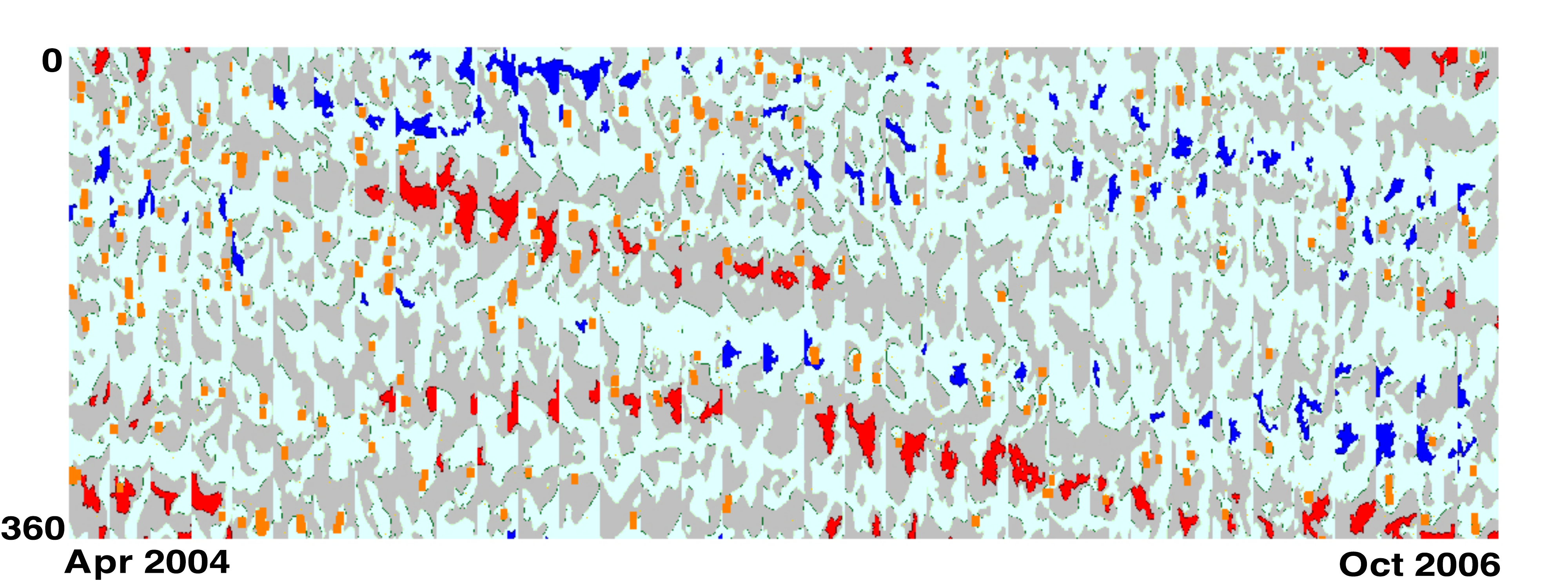} 
 \caption{Equatorial-zone stack plot as in Fig. \ref{figeq_mid}, CROT 2015 - 2049.}
   \label{figeq_midlate}
\end{center}
\end{figure}

\begin{figure}[b!]
\begin{center}
 \includegraphics[width=5.5in]{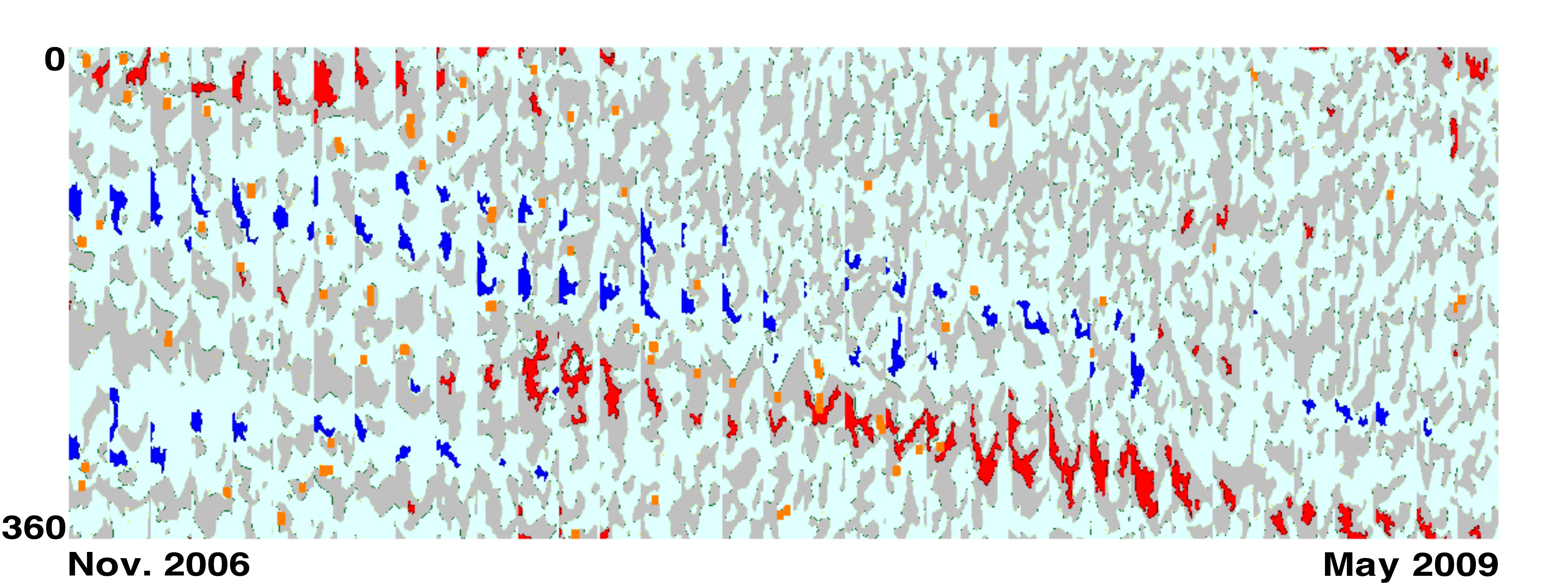} 
 \caption{Equatorial-zone stack plot as in Fig. \ref{figeq_mid}, CROT 2050 - 2084.}
   \label{figeq_late}
\end{center}
\end{figure}

\begin{figure}[b!]
\begin{center}
 \includegraphics[width=5.3in,height=6.in]{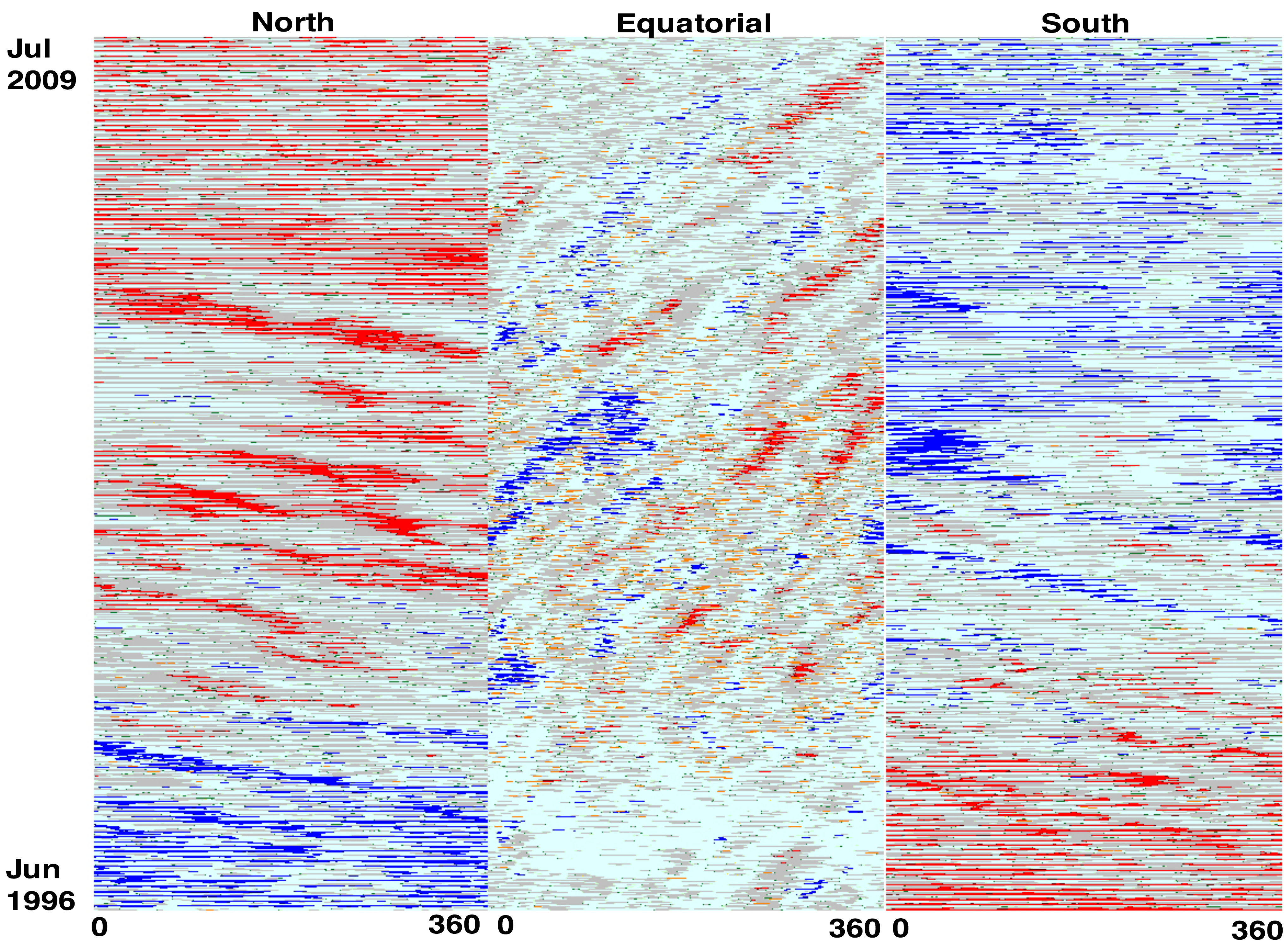} 
 \caption{Stack plots for (left) North polar zone ($30^\circ$ to $70^\circ$) latitude), (middle) Equatorial zone ($-20^\circ$ to $+20^\circ$ latitude), and (right) South polar zone ($-70^\circ$ to $-30^\circ$ latitude), for all of SC 23.  These are rotated $90^\circ$ in comparison to Figs. \ref{figeq_mid}-\ref{figeq_late}, so the horizontal axes = longitude.}
   \label{figallstack}
\end{center}
\end{figure}

Some of the original hand-drawn McIntosh maps were published as Upper Atmosphere Geophysics (UAG) reports in \cite[McIntosh (1975)]{mcintosh_75}, \cite[McIntosh and Nolte (1975)]{mcintoshnolte_75}, and \cite[McIntosh (1979)]{mcintosh_79}. Versions of the maps were also routinely published in the Solar-Geophysical Data (SGD) Bulletins. The UAG and SGD reports are all archived in scanned format at the NOAA National Center for Environmental Information (NCEI). However, many maps only existed in hard-copy format in boxes, and none of the scanned maps possessed metadata allowing digital search and analysis.  

The intent of the McA project has been first and foremost to preserve the archive in its entirety, by completing the scanning of all of the maps; this has been achieved. Second, we have designed and implemented a procedure to standardize the size and orientation of the digital maps, to remove any unnecessary notes, marks or symbols, and to colorize the maps so that each magnetic feature is uniquely searchable.  Fig. \ref{figcolormap} shows an example of a McA map processed in this manner.  To date, we have completed the full processing of solar cycle 23 (SC 23), and are working our way backwards through the earlier cycles.

\section{SC 23 Analysis}

We now present preliminary results for SC 23, and describe directions for future study.

{\underline{\it Active Longitudes and Periodic Solar-Wind Forcing}}.
Figs. \ref{figeq_mid} - \ref{figeq_late} show ``stack plots'' in which equatorial slices of maps such as Fig. \ref{figcolormap} are placed side by side so that trends over multiple Carrington rotations (CROTs) can be easily examined (see, e.g., \cite[McIntosh \& Wilson (1985)]{mcintoshwilson_85}). Stack plots are useful for illustrating active longitudes, where sunspots and magnetic flux appear to emerge preferentially (e.g., \cite[de Toma et al. (2000)])).  In particular, evidence for the persistence of two active longitudes separated by $180^\circ$ for 100 years has been found in historical records of sunspot locations (e.g., \cite[Berdyugina \& Usoskin (2003)]{berdus_03}) and in recently-digitized white-light images from the Kodaikanal Observatory (\cite[Mandal et al., 2016]{mandaletal_2016}).  

Figs. \ref{figeq_mid} - \ref{figeq_late} show not only sunspots (orange) but also near-equatorial coronal holes (red/blue). A $180^\circ$ longitudinal asymmetry is particularly evident during solar maximum years (Fig. \ref{figeq_mid}), and in general there is a long-lived pattern where coronal holes appear, disappear, and reappear in a preferred (differentially-rotating) set of longitudes. The pattern is also apparent in the quiet sun polarity (light-blue/grey) as noted previously in, e.g., \cite[McIntosh \& Wilson (1985)]{mcintoshwilson_85}. (See also \cite[Bilenko \& Tavastsherna (2016)]{biltav_16} for a study of coronal hole latitudinal and longitudinal patterns spanning three solar cycles.)  

An association of sunspot active longitudes with high-speed solar wind periodicities has been noted for some time (e.g., \cite[Balthasar \& Schuessler (1983)]{baltschu_83} and references therein).  The role of long-lived, low-latitude coronal holes in driving periodic behavior, both in the solar wind and in the Earth's space environment and upper atmosphere, was studied extensively for the SC 23 declining period and the extended solar minimum that followed (see, e.g., \cite[Temmer et al. (2007)]{temmeretal_07}, \cite[Gibson et al. (2009)]{gibetal_09}, and \cite[Luhmann et al. (2009)]{luhmannetal_09}).  As Fig. \ref{figeq_late} demonstrates, this otherwise quiet (low sunspot activity) time period was characterized by remarkably sustained longitudinal structure, first with three and then two near-equatorial open-field regions.

{\underline{\it Differential Rotation}}.  The downward shift of the active longitude locations with time seen in Figs. \ref{figeq_mid} - \ref{figeq_late}  corresponds to an increase in Carrington longitude and so indicates a rotation that is faster than the Carrington rate. This is expected based on solar surface differential rotation, since the Carrington rate (27.2753 days as viewed from the Earth) corresponds to a mid-latitude surface rotation rate.  Fig. \ref{figallstack} shows an extended stack plot for all of SC 23, both for the equatorial zone shown in Figs. \ref{figeq_mid} - \ref{figeq_late} and for the northern and southern polar zones.  Near solar minimum (the top and bottom of the plots) the polar zones show unipolar coronal holes at all latitudes, but at solar maximum and in the declining phase the slower polar rotation rate is evident. Note that Fig. \ref{figallstack} is rotated $90^\circ$ relative to Figs. \ref{figeq_mid} - \ref{figeq_late}, so that a feature moving at the Carrington rate would appear vertical (constant Carrington longitude).

Studies of sunspot active-longitude differential rotation (e.g., \cite[Usoskin et al. (2005)]{usoskinetal_05}, \cite[Mandal et al. (2016)]{mandaletal_2016}) find a rotation rate of the active longitude location matching the surface flow at that latitude. This is in contrast to individual sunspots which generally rotate faster than the surface does, indicating they are rooted below the near-surface shear layer (\cite[Thompson et al. (2003)]{thompsonetal_03}).  Studies of coronal hole rotation indicate that polar coronal holes may rotate rigidly, while low-latitude coronal holes rotate more differentially but with a variability possibly associated with coronal-hole lifetime and solar-cycle phase (see, e.g., \cite[Ikhsanov \& Ivanov (1999)]{ikhiv_99}).  Such variability may contribute to the range in slopes of coronal-hole patterns seen in Figs. \ref{figeq_mid} - \ref{figallstack}.

\begin{figure}[ht!]
\begin{center}
\includegraphics[width=3.35in]{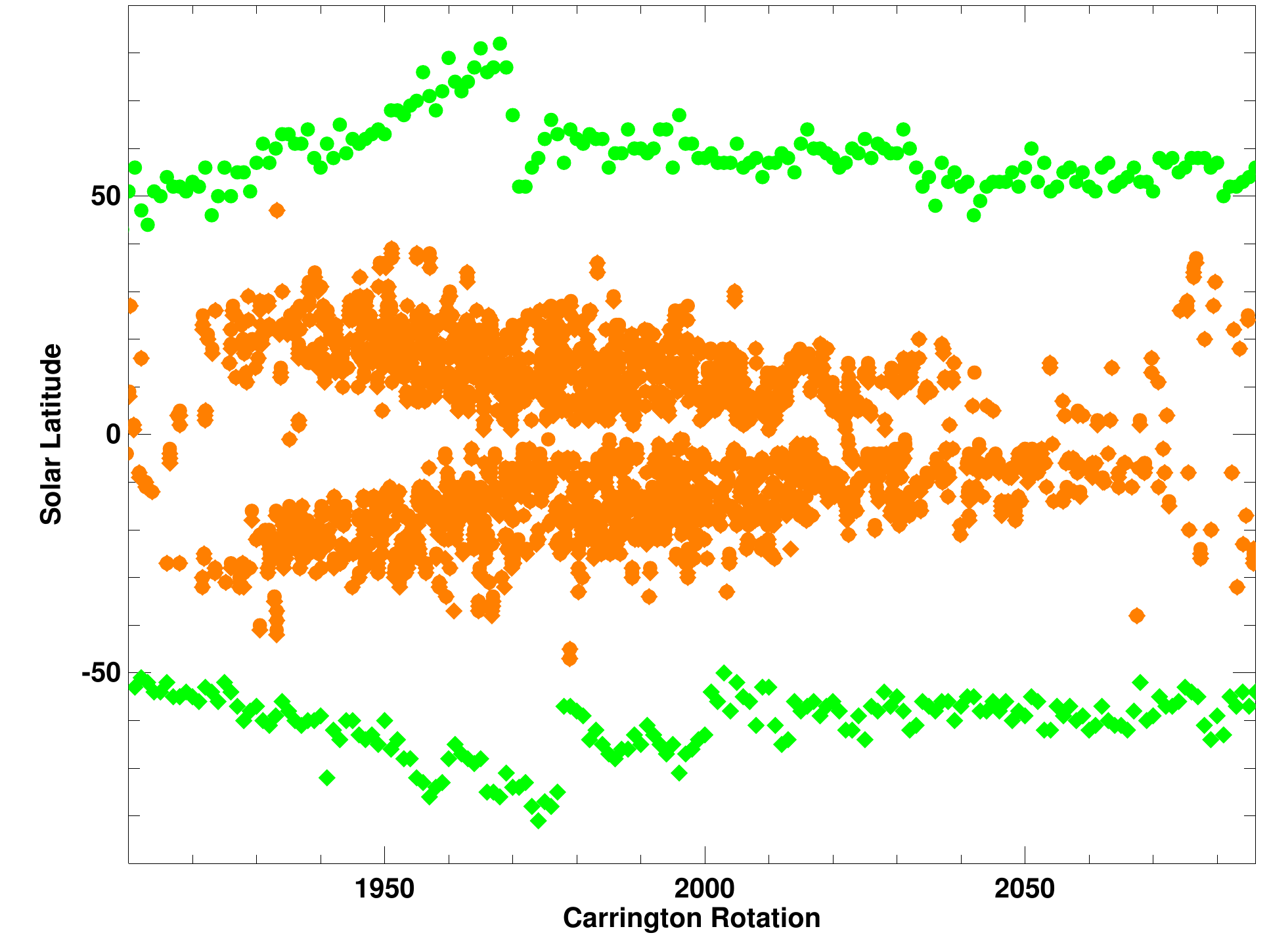} 
\includegraphics[width=3.3in]{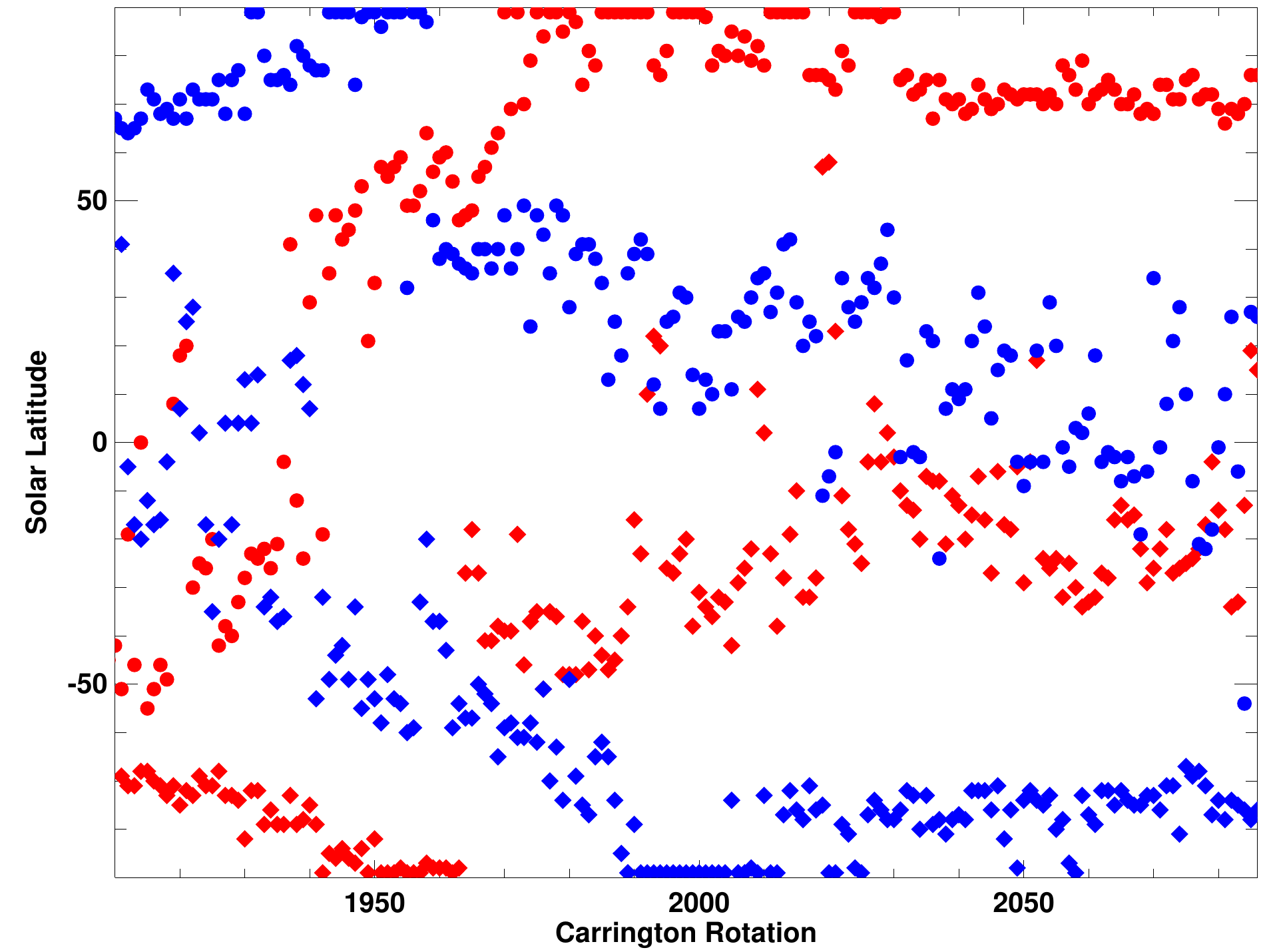} 
\includegraphics[width=3.3in]{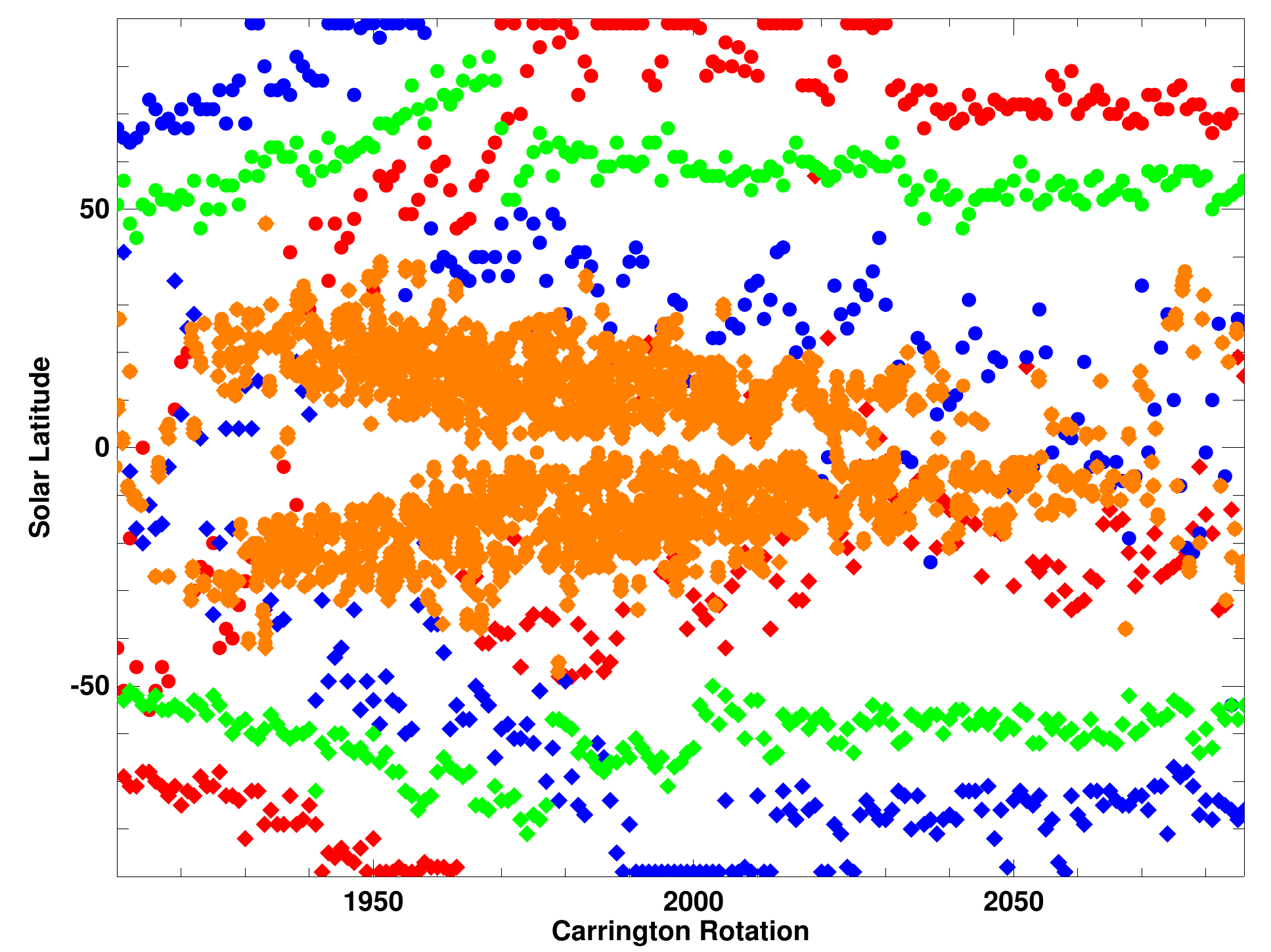} 
 \caption{SC 23 patterns of open vs. closed magnetic features. (Top) sunspots (orange) and most poleward filament (green). (Middle) coronal hole boundaries (red=negative, blue=positive; furthest north per CROT=circles, furthest south=diamonds). (Bottom) combined.}
   \label{figbutterfly}
\end{center}
\end{figure}

{\underline{\it Evolution of Open vs. Closed Fields}}.  Fig. \ref{figbutterfly} shows the global evolution of coronal holes, sunspots, and filaments over SC 23.  The sunspots are plotted as a classic butterfly diagram, appearing after solar minimum first at high latitudes, and emerging progressively closer to the equator as the cycle continues.  The green dots show the location of the most poleward filament for each CROT, and show the ``rush to the poles'' which is a tracer of the  solar magnetic polarity reversal process. After this reversal, the old polar crown filament is replaced by a secondary crown filament in both hemispheres (\cite[McIntosh (1992)]{mcintosh_92}).  

The middle panel in Fig. \ref{figbutterfly} plots the locations of coronal hole boundaries as a function of CROT. In particular, the northmost and southmost boundary position is plotted for both magnetic polarities, capturing the behavior of both polar and low-latitude coronal holes.  The result is the ``double helix'' pattern referred to in \cite[S. McIntosh et al. (2014)]{mcintoshetal_14} (see also \cite[Bilenko \& Tavastsherna (2016)]{biltav_16}, \cite[Fujiki et al. (2016)]{fujiki_16}).  This process of polar coronal holes reforming as a consequence of the migration of lower-latitude coronal holes poleward is described in \cite[Webb et al. (1984)]{webbetal_84} and \cite[Harvey \& Recely (2002)]{harveyrecely_02}.  Finally, the bottom panel in Fig. \ref{figbutterfly} shows how the positive and negative polarity open fields (blue and red) surround the closed-field filaments (green) and sunspots (orange), on a global scale.

\section{Conclusions}

The unique power of the McIntosh archive is its capability for simultaneously representing closed and open magnetic structures over a range of time scales. The completion of the full McA digitization will provide the community with a comprehensive resource for addressing key questions including: How do active longitudes vary within and between solar cycles, for both closed and open magnetic features? Where are closed and open magnetic features rooted (as evidenced by rotation rate), and how does this depend on solar cycle phase, feature lifetime, and latitude? How does the evolution of open and closed magnetic features relate to surface flows on solar-cycle time scales (e.g., torsional oscillations, \cite[Howe et al., 2013]{howeetal_13})? Answering any or all of these questions has important implications for our understanding of the solar dynamo, and for our interpretation of periodic variations of Earth's space environment and upper atmosphere.

\acknowledgements{We dedicate this project to Pat McIntosh, who died this year. We are grateful to his daughter Beth Schmidt for granting us permission to use his original data. The work of the authors is supported by NSF RAPID grant number 1540544. NCAR is funded by the NSF.  We thank Giuliana de Toma, Larisza Krista, and Scott McIntosh for helpful discussions.}

\end{document}